\documentclass[a4paper,11pt]{article}
\usepackage{pos}

\title{ Exotic states in the $\pi D$
system}
\ShortTitle{Exotics in the $\pi D$ system}

\author*[a,b]{Eric B.~Gregory}
\author[c,d]{Feng-Kun Guo}
\author[a,e,f]{Christoph Hanhart}
\author[a,b,f]{Stefan Krieg}
\author[a,e,f]{Thomas Luu}

\affiliation[a]{Institute for
           Advanced Simulation, 
           Forschungszentrum J\"ulich, D-52425 J\"ulich, Germany}
\affiliation[b]{JARA-HPC, Jülich Supercomputing Center, Forschungszentrum Jülich, 54245 Jülich Germany}
\affiliation[c]{CAS Key Laboratory of Theoretical Physics,
            Institute of Theoretical Physics,\\ Chinese Academy of Sciences,
            Beijing 100190, China}
\affiliation[d]{School of Physical Sciences,
            University of Chinese Academy of Sciences,
            Beijing 100049, China}
\affiliation[e]{Institut f\"ur Kernphysik and
            J\"ulich Center for Hadron Physics,\\ 
            Forschungszentrum J\"ulich, D-52425 J\"ulich, Germany}
\affiliation[f]{Helmholtz-Institut f\"ur Strahlen- und
            Kernphysik and Bethe Center for Theoretical Physics,\\
            Universit\"at Bonn,  D-53115 Bonn, Germany}

\emailAdd{e.gregory@fz-juelich.de}

\abstract{

In this proceedings we consider several states, namely the  $D^*_{s0}(2317)$, $D_{s1}(2460)$, $D^*_{0}(2300)$ and $D_{1}(2430)$, which appear to defy description as simple quark-antiquark pairs. Theoretical input from unitarized chiral perturbation theory suggests they can be understood as 
emerging from Goldstone-Boson--$D$-meson scattering.

We present results from an $SU(3)$ flavor-symmetric lattice QCD simulation at large pion masses suggesting that there  exists a $\pi D$ bound state in the flavor-sextet representation that cannot emerge for quark-antiquark states, but that appears naturally from the multiquark states.
Moreover, we find repulsion in the [15] representation, which establishes the pattern predicted for the interactions of Goldstone bosons with $D$ mesons. This suggests these states may have the structure of hadronic molecules.

}

\FullConference{%
 The 38th International Symposium on Lattice Field Theory, LATTICE2021
  26th-30th July, 2021
  Zoom/Gather@Massachusetts Institute of Technology
}

\begin{document}
\maketitle

\section{Introduction}

A major goal of  modern particle physics is to explain the nature of particles observed in experiments. We focus on several observed particles, namely $D^*_{s0}(2317)$, $D_{s1}(2460)$, $D^*_{0}(2300)$ and $D_{1}(2430)$, which defy explanation through the quark model as simple $q\overline{q}$ meson states. The $D^*_{s0}(2317)$ and  $D_{s1}(2460)$ states are lighter than quark model predictions would suggest~\cite{Godfrey:1985xj,Godfrey:2015dva,Ebert:2009ua}. Furthermore, we expect that non-strange states should be about 150~MeV lighter than their strange counterparts. However, $D^*_{0}(2300)$ and $D_{1}(2430)$ are  similar in mass to their corresponding strange states.

The likely explanation is that these states are exotic. Insight into their nature has come from unitarized chiral perturbation theory  (U$\chi$PT)~\cite{Kolomeitsev:2003ac,Albaladejo:2016lbb,Du:2017zvv,Liu:2012zya, Guo:2018kno}. U$\chi$PT suggests that $D^*_{s0}(2317)$ is a $DK$ molecule and that the very broad $D^*_{0}(2300)$ state contains two poles around 2105~MeV and 2451~MeV. Also the $D^*_{s0}(2317)$ is the $SU(3)$-flavor partner of the lower of these two poles, while the upper pole is a member of an $SU(3)$-sextet.

Several lattice studies have added information about these states, with \cite{Mohler:2013rwa} examining $DK$ scattering. The Regensburg group calculated $D^*_{s0}(2317)$ and $D_{s1}(2460)$ masses~\cite{Bali:2017pdv}. The Hadron Spectrum Collaboration predicted that the lightest $D^*_0$ should be below 2300MeV. For reviews, see Refs.~\cite{Guo:2017jvc,Guo:2019dpg}.

There remain, however, questions regarding the structure of these states: are they compact tetraquarks of the diquark-antidiquark type or are they hadronic molecules, as suggested by U$\chi$PT? 

Further insight comes from understanding the nature of the $SU(3)$-flavor multiplets. In a flavor-symmetric world, a four-quark state $c\overline{q}q\overline{q}$ breaks down as
\begin{equation}
    [\bar{3}]\otimes [8]
=[15]\oplus[6]\oplus[\bar{3}]\ ,
\end{equation}
ignoring the $[\overline{3}]$ arising from $c\overline{q}$ coupling to the flavor-singlet $q\overline{q}$. Dmitrasinovic predicted that in a tetraquark configuration, all multiplets are attractive~\cite{Dmitrasinovic:2005gc}. On the other hand, U$\chi$PT predicts that for hadronic molecules, the $[\overline{3}]$ would be the most attractive multiplet, followed by the $[6]$, with the $[15]$ being repulsive~\cite{Kolomeitsev:2003ac,Guo:2006fu,Guo:2009ct}.

With these conflicting predictions, determination of the attractive or repulsive nature of the $SU(3)$ multiplets could lend strong support to either the compact tetraquark or hardonic molecule hypothesis. 
With this in mind, we set out to do a $SU(3)$-flavor symmetric  lattice calculation of the $D\pi$ system.

\section{Lattice Calculation}
\subsection{Ensemble tuning}
The U$\chi$PT calculation of Du {\em et al.}~\cite{Du:2017zvv} suggests that $SU(3)$ flavor symmetric light quarks tuned so that the pion mass $M_\pi$ roughly in the range 600 -- 700~MeV should produce a near-threshold virtual or bound state in the$[6]$ representation. Our goal then was to generate ensembles with three light quarks tuned to give the flavor symmetric pions in this mass range, and the charm quark at its physical mass.  

Using CHROMA~\cite{chroma_ref}, with either QPHIX\cite{qphix_ref} or QUDA\cite{quda_ref}, we generated 63 tuning ensembles with  clover Wilson fermions with six iterations of stout smearing. We initially explored three values of the coupling, $\beta=3.4,3.5$ and $3.6$.

 The tuning procedure is complicated by the lack of light quarks near the physical mass range. This prevents us from setting the lattice scale with traditional light hadron observables and methods such as renormalization flow, which are calibrated with the light hadron spectrum. Instead we use a three-step tuning procedure and rely on charmonium splitting to set the lattice scale.
 
\begin{figure}[ht]
     \centering
     \includegraphics[width=0.8\textwidth]{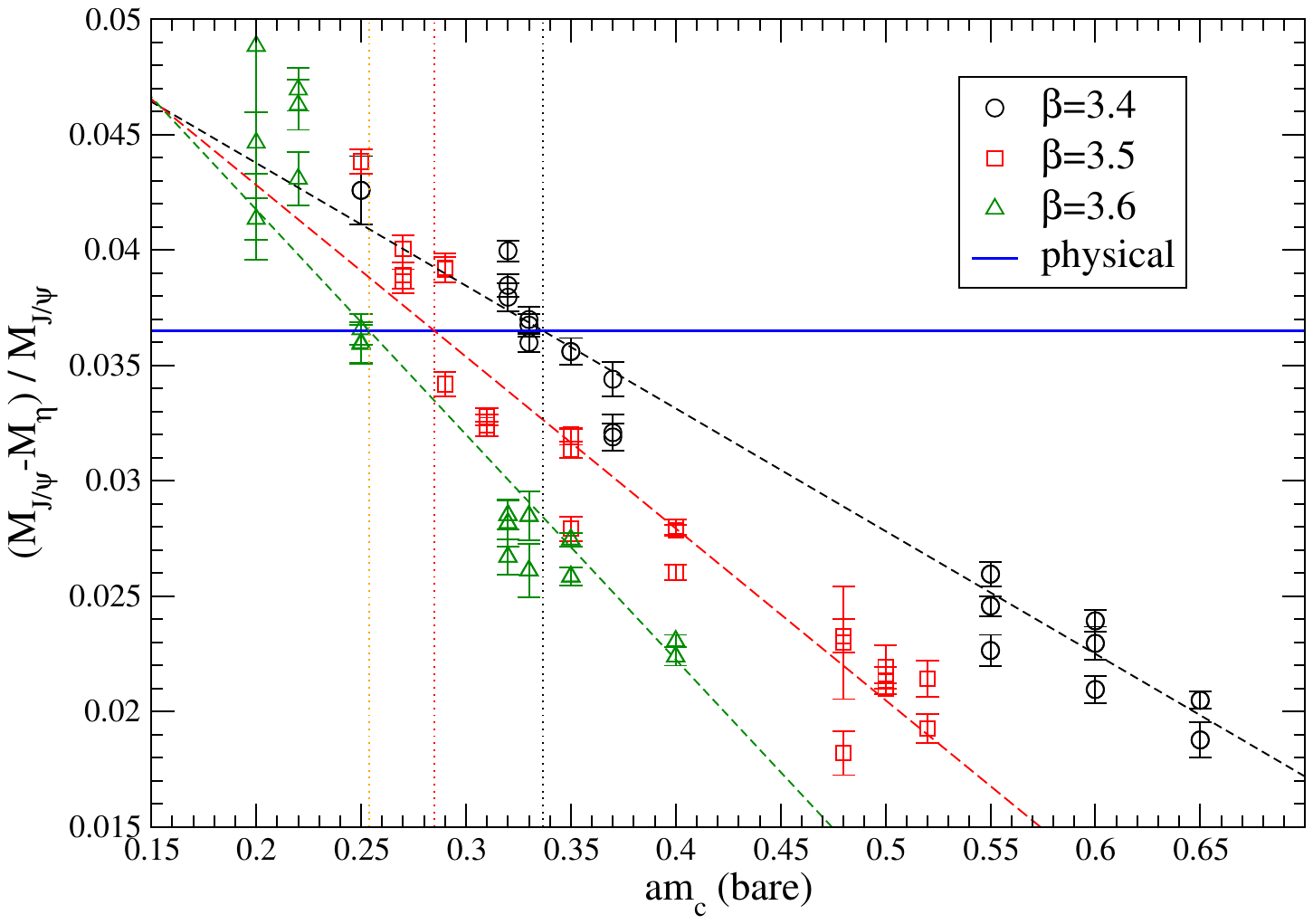}
     \caption{Tuning $m_c$. For each $\beta$, we find the value of $am_c$ where the linear fit to measured values of the charmonium splitting ratio (dashed lines) crosses the physical value $R=0.0365$ (blue line).}
     \label{fig:split_Jpsi_ratio}
 \end{figure}

 In the first step we construct the ratio
 \begin{equation}
    R= \frac{M_{J/\psi}-M_{\eta_c}}{M_{J/\psi}}
 \end{equation}
 and vary the bare charm quark mass until the physical value of $R=0.365 $ is achieved. We then read off the value of $am_c$ that gives physical charm quarks. This step is illustrated in Fig.~\ref{fig:split_Jpsi_ratio}.
 
\begin{figure}[ht]
     \centering
     \includegraphics[width=0.8\textwidth]{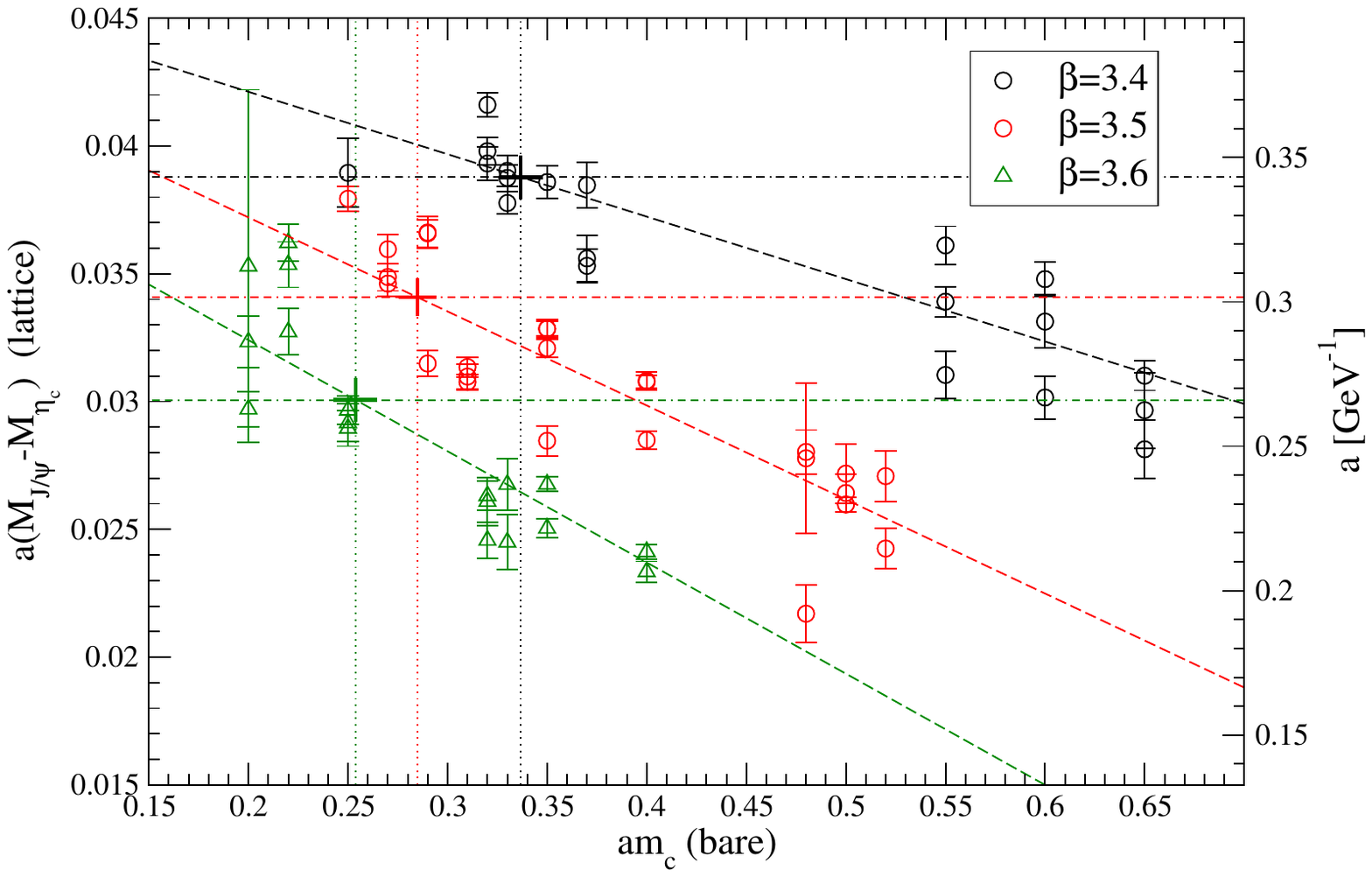}
     \caption{Determining the lattice scale $a$. Where the linear fits to lattice values of the splitting $a(M_{J/\psi}-M_{\eta_c})$ intersect the value of $am_c$ determined for each $\beta$ in the first step, we use this splitting to determine the lattice spacing $a$ in GeV$^{-1}$ (horizontal dashed lines to the right axis).}
     \label{fig:split_Jpsi_etac_spac}
 \end{figure}

Once we have determined the target $am_c$ for each $\beta$, we use the value of the charmonium splitting  $aM_{J/\psi}-aM_{\eta_c}$ to establish the lattice spacing:
\begin{equation}\label{eq:lat_spac}
a=\frac{\left(aM_{J/\psi}-aM_{\eta_c}\right)_{\rm latt}}{\left(M_{J/\psi}-M_{\eta_c}\right)_{\rm phys}} = \frac{\left(aM_{J/\psi}-aM_{\eta_c}\right)_{\rm latt}}{113 ~{\rm MeV}}.
\end{equation}
This process is illustrated in Fig.~\ref{fig:split_Jpsi_etac_spac}.

 We note that we do not include disconnected diagrams in our calculation of $M_{J/\psi}$ and $M_{\eta_c}$ which we estimate introduces a 10\% systematic overestimation of $\left(aM_{J/\psi}-aM_{\eta_c}\right)_{\rm latt}$ and hence a corresponding error in $a$ \cite{Hatton:2020qhk}.
 
\begin{figure}[ht]
     \centering
     \includegraphics[width=0.8\textwidth]{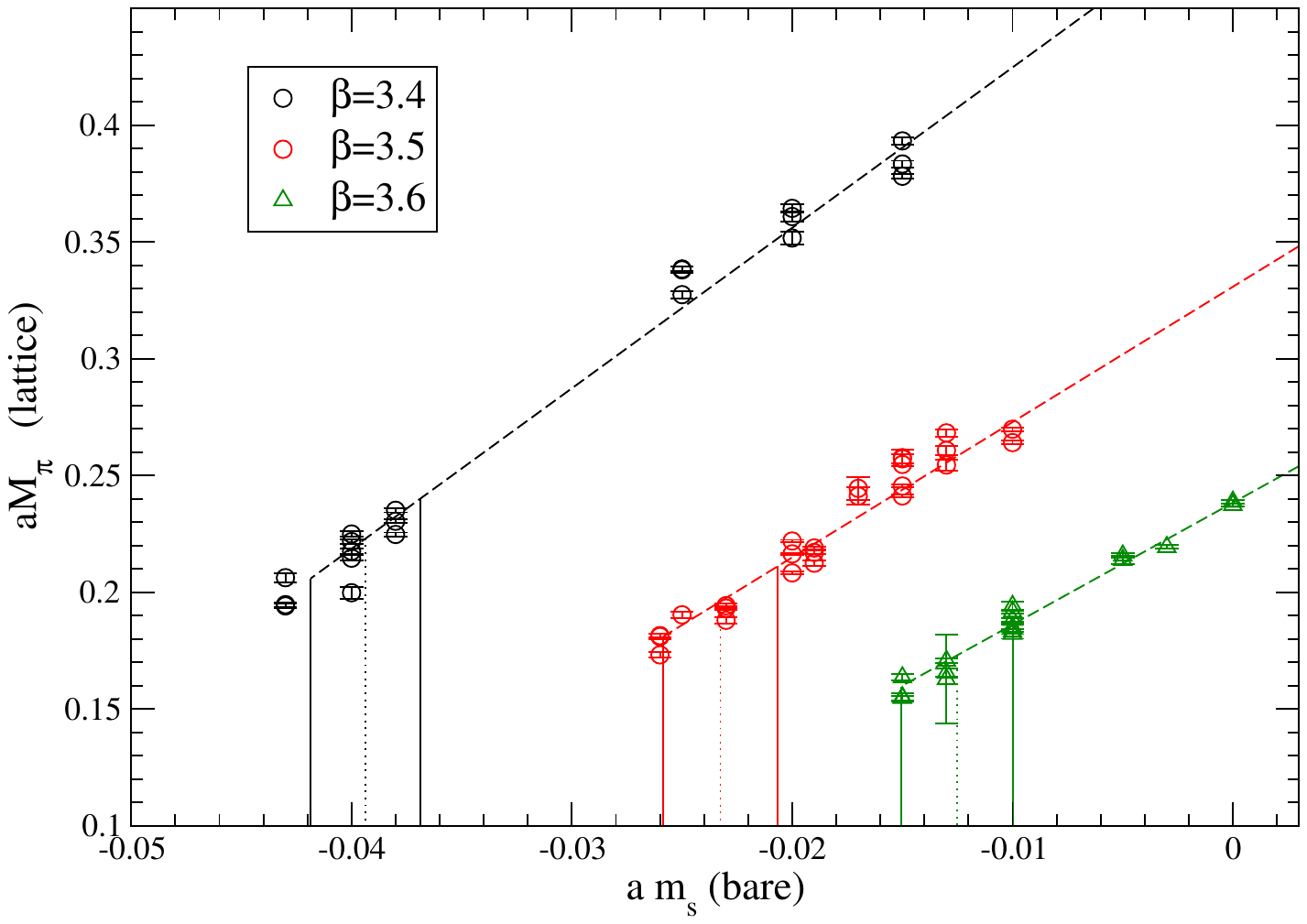}
     \caption{Tuning the light ($SU(3)$) quark masses. The dashed diagonal lines  are linear fits to ($m_q$,$aM_{\pi}$). The dashed vertical lines drop from the point the fit lines pass through $a(650~{\rm MeV})$. The vertical solid lines correspond to 600~MeV and 700~MeV. Different $a$ for each $\beta$ is implied.}
     \label{fig:Mps_msbare}
 \end{figure}

The final step is to find the value of the bare light quark mass, $am_s$, that produces pions in the 600 -- 700~MeV  range. This is straightforward now that we have an estimate for the lattice spacing $a$. See Fig.~\ref{fig:Mps_msbare}.

We reiterate that, with the exception of the charm quark, we are not tuning to physical quark masses, nor are we attempting precision spectroscopy requiring a continuum extrapolation. For our purposes a single lattice spacing is sufficient to investigate the nature of the $SU(3)$ multiplets. For this reason, we focus from here on the $\beta=3.6$ ensembles.

We find that for $\beta=3.6$, $a=0.27(2)_{\rm stat}(2)_{\rm sys}$ GeV$^{-1}$. At this lattice spacing, our ensemble at $am_q=-0.013$, $am_c=0.25$ is very close to the target point and corresponds to $M_\pi=612(90)$~MeV (with the uncertainty dominated by the lattice spacing determination).  At this target point we generated three ensembles, each with $L_t=64$ lattice units in the time direction, but varying spatial volumes of $L_s^3=32^3, 40^3$ and $48^3$. These correspond  roughly to spatial sizes of 1.6, 2.1 and 2.6 fm, respectively. For each we generated 2500 -- 2700 trajectories.

\subsection{Interpolating operator design and measurement}

Because the $[\overline{3}]$ is not useful for distinguishing between tetraquarks and hadronic molecules, and because of the additional computational difficulty in resolving its disconnected diagrams, we do not address the $[\overline{3}]$ irrep in the lattice simulation described below. We focus instead on the $[6]$ and $[15]$.

To construct the appropriate interpolating operators for the $[6]$ and $[15]$ we  first construct the needed $SU(3)$ flavor states within a tensor basis~\cite{Georgi:1982jb}.  Since the $c$ quark is in an $SU(3)$ singlet, the states for the remaining degenerate light quarks can be constructed with appropriate projection onto the $[6]$ and $[15]$ irreps.  Table \ref{tab:6} shows the states and their quantum numbers for the $[6]$ irrep.

\begin{table}
\center
\caption{The light quark content of the states in the $[6]$ representation and their associated quantum numbers.  $T^2_a$ is the Casimir operator, $I_z$ is the third component of isospin, and $Y$ is the hypercharge.\label{tab:6}}
\begin{tabular}{c|c|c|c|c}
state&components& $T_a^2$ & $I_z$ & $Y$\\
\hline\hline
1& $-|u\bar{u}\bar{d}\rangle\frac{1}{2}+|u\bar{d}\bar{u}\rangle\frac{1}{2}-|s\bar{d}\bar{s}\rangle\frac{1}{2}+|s\bar{s}\bar{d}\rangle\frac{1}{2}$ &$\frac{10}{3}$ &$+\frac{1}{2}$ & $-\frac{1}{3}$\\
2& $|d\bar{u}\bar{d}\rangle\frac{1}{2}-|d\bar{d}\bar{u}\rangle\frac{1}{2}-|s\bar{u}\bar{s}\rangle\frac{1}{2}+|s\bar{s}\bar{u}\rangle\frac{1}{2}$ &$\frac{10}{3}$ &$-\frac{1}{2}$ & $-\frac{1}{3}$\\
\hline
3& $|u\bar{u}\bar{s}\rangle\frac{1}{2}-|u\bar{s}\bar{u}\rangle\frac{1}{2}-|d\bar{d}\bar{s}\rangle\frac{1}{2}+|d\bar{s}\bar{d}\rangle\frac{1}{2}$ &$\frac{10}{3}$ &0 & $+\frac{2}{3}$\\
4& $|d\bar{s}\bar{u}\rangle\frac{1}{\sqrt{2}}-|d\bar{u}\bar{s}\rangle\frac{1}{\sqrt{2}}$ &$\frac{10}{3}$ &$-1$ & $+\frac{2}{3}$\\
5& $|u\bar{s}\bar{d}\rangle\frac{1}{\sqrt{2}}-|u\bar{d}\bar{s}\rangle\frac{1}{\sqrt{2}}$ &$\frac{10}{3}$ &$+1$ & $+\frac{2}{3}$\\
\hline
6& $|s\bar{d}\bar{u}\rangle\frac{1}{\sqrt{2}}-|s\bar{u}\bar{d}\rangle\frac{1}{\sqrt{2}}$ &$\frac{10}{3}$ &0 & $-\frac{4}{3}$\\
\end{tabular}
\end{table}
If we choose state 5, we would write down the operator:
\begin{equation}
O^5_{[6]}(x';x)=\frac{1}{\sqrt{2}}\left\{\left[\bar{s}(x')\Gamma c(x')\right]\left[\bar{d}(x)\Gamma u(x)\right]-
\left[\bar{d}(x')\Gamma c(x')\right]\left[\bar{s}(x)\Gamma u(x)\right]\right\}.
\end{equation}
We could instead write the operator for state 6:
\begin{equation}
O^5_{[6]}(x';x)=\frac{1}{\sqrt{2}}\left\{\left[\bar{s}(x')\Gamma c(x')\right]\left[\bar{d}(x)\Gamma u(x)\right]-
\left[\bar{d}(x')\Gamma c(x')\right]\left[\bar{s}(x)\Gamma u(x)\right]\right\}.
\end{equation}
Regardless, at the SU(3) point we get the same expression when we complete the contractions to obtain the correlator:
\begin{multline}\label{eqn:6contraction}
\langle O^{i}_{[6]}(y';y)\bar{O}^{i}_{[6]}(x;x)\rangle = 
\text{Tr}\left[\Gamma \gamma_5\mathcal{S}^\dag_{y';x}\gamma_5\Gamma\mathcal{S}^{}_{y';x}\right]
\text{Tr}\left[\Gamma \gamma_5\mathcal{S}^\dag_{y;x}\gamma_5\Gamma\mathcal{C}^{}_{y;x}\right]\\
+\text{Tr}\left[\Gamma \gamma_5\mathcal{S}^\dag_{y;x}\gamma_5\Gamma\mathcal{S}^{}_{y';x}\Gamma \gamma_5\mathcal{S}^\dag_{y';x}\gamma_5\Gamma\mathcal{C}^{}_{y;x}\right]\ ,
\end{multline}
The contraction for the $[15]$, $\langle O^{i}_{[d]}(y';y)\bar{O}^{i}_{[d]}(x;x)\rangle^{[15]} $ is identical to Eq.~\eqref{eqn:6contraction}, but with a relative minus sign between the terms.  In our calculations we consider only $\Gamma=\gamma_5$.

\subsection{Analysis}

\begin{figure}
    \centering
    \vspace{-0.2in}
    \includegraphics[width=0.8\columnwidth]{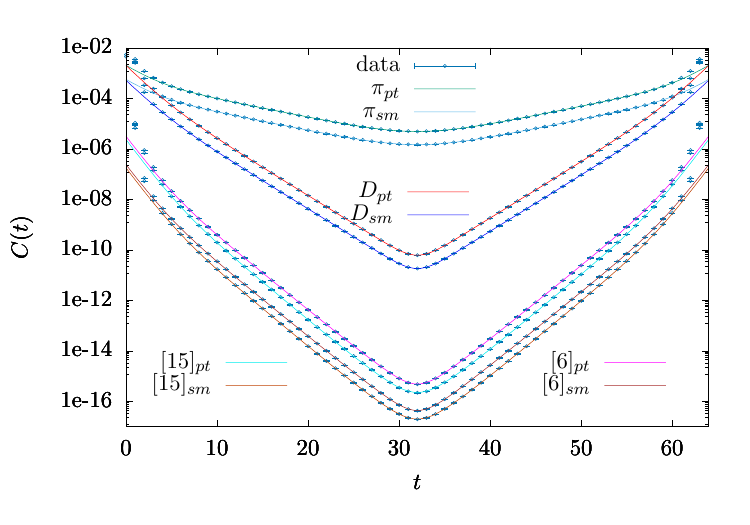}
    \caption{Eight hadron correlators used for simultaneous 2-state fits with $t_{\rm min}=5$ (see text). From top to bottom, the correlators represent: $\pi_{\rm pt}$, $\pi_{\rm sm}$, $D_{\rm pt}$, $D_{\rm sm}$, $[6]_{\rm pt}$, $[15]_{\rm pt}$, $[6]_{\rm sm}$, and $[15]_{\rm sm}$, with the subscripts ``pt" and ``sm" referring to point and smeared sink operators. Lattice data error bars are too small to be seen.}
    \label{fig:all_corrs}
\end{figure}
We measure correlators for the $\pi$, $D$, as well as the $[6]$ and $[15]$ states, using point and smeared sink operators for each. This gives us eight different correlators, which we fit simultaneously using
\begin{equation}
    C_{P,s}(t) =\sum_{j=0}^{(N-1)}
    A_{P,s,j}\cosh\left(M_{P,j}(t-L_t/2)\right)
\end{equation}
for $P=\{\pi, D\}$ and 
\begin{multline}\label{eq:fitform}
    C_{P,s}(t) =
    B_{s}\cosh\left((M_D-M_\pi)(t-L_t/2)\right)
    +A_{P,s,0}\cosh\left((\Delta_{M_P}+M_{D,0}+M_{\pi,0})(t-L_t/2)\right)\\
    + \sum_{j=1}^{(N-1)} A_{P,s,j}\cosh\left(M_{p,j}(t-L_t/2)\right),
\end{multline}
for $p=\{ [6], [15]\}$, and $s=$\{point, smeared\} sink operators.
The overall fit form is
\begin{equation}
{\mathcal C}(t,P,s) = \delta_{P,P'}\delta_{s,s'}C_{P',s'}(t).
\end{equation}
The first term in Eq.~\eqref{eq:fitform}  is due to forward-propagating $\pi$ and backwards-propagating $D$ and vice versa. The parameters of interest are 
\begin{eqnarray}
    \Delta M_{[6]}&\equiv& M_{[6]}-\left(M_D+M_\pi\right), \nonumber\\
    \Delta M_{[15]}&\equiv& M_{[15]}-\left(M_D+M_\pi\right).
\end{eqnarray} These will be negative for attractive states and positive for repulsive scattering states. The correlators and sample fits are shown in Fig.~\ref{fig:all_corrs}.

We performed ground-state ($N=1$) and ground-state plus first-excited-state ($N=2$) fits, varying a symmetric fit range $t_{\rm min}...(L_t-t_{\rm min})$.
We used jackknife resampling with a bin size of 15 (150 trajectories) to extract uncertainties on the fit parameters.

\section{Results, conclusions and future work}

\begin{figure}
    \centering
    \vspace{-0.7in}
    \includegraphics[width=\columnwidth]{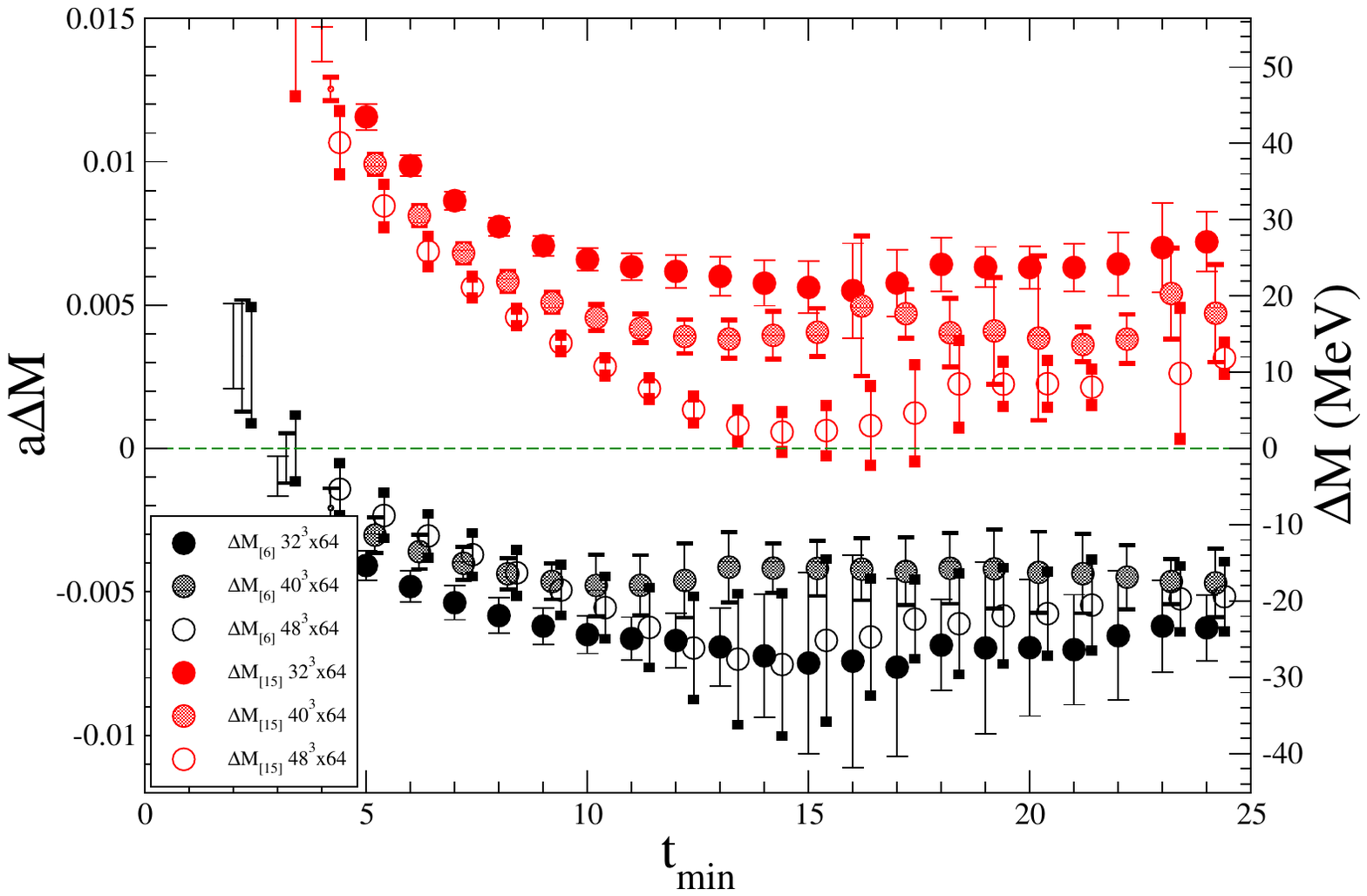}
    \caption{Mass shifts $\Delta{M_{[6]}}$ and $\Delta{M_{[15]}}$ from two-state fits  ($N=2$) as a function of $t_{\rm min}$. A visible symbol indicates that the fit confidence ($p$-value) is near 1. Error bars are larger on the $40^3$ fits because there are fewer measurements per configuration.}
    \label{fig:2state_fits}
\end{figure}

Figure~\ref{fig:2state_fits} shows the resulting $\Delta_M$ values.
We find a statistically significant negative signal for $\Delta_{M_{[6]}}$, with little variation across the three volumes.
For the $[15]$ we see $\Delta_{M_{[15]}}>0$ for all three volumes. However, $\Delta_{M_{[15]}}$ seems to approach zero as the volume increases, as would be expected in a scattering state.

Taken together these are consistent with the U$\chi$PT predictions of the  $SU(3)$ $[6]$ and $[15]$ multiplets, and thus support the interpretation of the lowest-lying positive parity charmed mesons  as hadronic molecules. In particular, the [6] state is beyond the simple $c\bar q$ configurations and is explicitly exotic.

Our next step in future work is to remove the remaining ambiguity about these results by increasing the statistics, using more advanced fitting techniques, and performing a L\"{u}scher finite-volume analysis.

\section{Acknowledgements}
We thank Alessandro Pilloni,  Evan Berkowitz, Andrea Shindler, David Wilson and Christopher Thomas for useful discussions. 
We are  is indebted to Andr\'{e} Walker-Loud for his insightful discussions related to backward propagating states. We are grateful to Balint Joo for assistance with CHROMA and to Kate Clark and Mathias Wagner for guidance with QUDA.
This work is supported in part by the National Natural Science Foundation of China (NSFC) under Grant No.~11835015,
No.~12047503, and No.~11961141012, by the NSFC and the Deutsche Forschungsgemeinschaft (DFG, German Research
Foundation) through the funds provided to the Sino-German Collaborative
Research Center ``Symmetries and the Emergence of Structure in QCD''
(NSFC Grant No.~12070131001, DFG Project-ID 196253076 -- TRR110), and by the Chinese Academy of Sciences (CAS) under Grant No.~XDB34030000
and No.~QYZDB-SSW-SYS013. 

The authors gratefully acknowledge the computing time granted by the JARA Vergabegremium and provided on the JARA Partition part of the supercomputer JURECA\cite{jureca} at Forschungszentrum J\"{u}lich, through project CJIAS0421, and the Gauss Centre for Supercomputing e.V. (www.gauss-centre.eu) for funding this project by providing computing time on the GCS Supercomputer JUWELS\cite{juwels} at J\"{u}lich Supercomputing Centre (JSC), through project HADRONSEXTCONT.

\end{document}